\documentclass[11pt]{article}

\usepackage[text={6.8in,8.5in},centering]{geometry}
\usepackage{times}
\usepackage{amsthm}
\usepackage{amsmath,amsfonts,amssymb} 
\usepackage{array,delarray,xspace} 
\usepackage[pdftex]{graphicx} 
\usepackage{dsfont} 
\usepackage{xspace} 
\usepackage{psfrag}
\usepackage{epsfig}
\usepackage{graphics}
\usepackage{epic}
\usepackage{eepic}
\usepackage{color}
\usepackage{wrapfig}
\usepackage{multirow}
\usepackage{url}
\usepackage{subfig}
\usepackage{hyperref}

\def\final{0} 
\ifnum\final=1  
\newcommand{\vnote}[1]{[{\small Vicky: \bf #1}]\marginpar{*}}
\newcommand{\sidecomment}[1]{\marginpar{\tiny #1}}
\else 
\newcommand{\vnote}[1]{}
\newcommand{\sidecomment}[1]{}
\fi  

\newtheorem{lemma}{Lemma}[section]
\newtheorem{thm}[lemma]{Theorem}
 \newtheorem{mdef}[lemma]{Definition}

\newcommand{\ms}[1]{\ensuremath{\mathsf{#1}}}
\newcommand{\bra}[1]{\ensuremath{\langle#1|}}
\newcommand{\ket}[1]{\ensuremath{|#1\rangle}}
\newcommand{\argmax}{\operatornamewithlimits{arg\ max}}

\def\reals{{\mathbb R}}

\renewcommand{\bar}{\overline}

\newcommand{\ver}{{\ms{V}}}
\newcommand{\edge}{{\ms{E}}}

\newcommand{\nbr}{{\ms{nbr}}}

\newcommand{\energy}{{\mathcal{E}}}
\newcommand{\eng}{{\mathcal{E}}}
\newcommand{\oy}{{\mathcal{Y}}}

\newcommand{\ham}{{\mathcal{H}}}

\newcommand{\EC}{Exact Cover }
\newcommand{\SEC}{{\sc{EC3 } }}

\newcommand{\SAT}{3SAT }

\newcommand{\gmin}{g_{\ms{min}}}
\newcommand{\wmis}{{\ms{mis}}}

\newcommand{\GEC}{G_{\ms{EC}}}
\newcommand{\GSAT}{G_{\ms{SAT}}}

\begin{document}

\title{Different Adiabatic Quantum Optimization Algorithms for the NP-Complete Exact Cover and 3SAT Problems}
\author{Vicky Choi 
\\{\em vchoi@cs.vt.edu}
\\ Department of Computer Science
\\ Virginia Tech
\\Falls Church, VA
 }
\maketitle

\begin{abstract}
One of the most important questions in studying quantum computation
is: whether a quantum computer can
 solve NP-complete problems  
more efficiently than a classical computer? 
In 2000, Farhi, et al. (Science,
292(5516):472--476, 2001) proposed the adiabatic quantum optimization (AQO), a paradigm that
directly attacks NP-hard optimization
problems.
How powerful is AQO? Early on, van Dam 
and Vazirani claimed that AQO failed (i.e. would take exponential
time) for a family of 3SAT instances they constructed.
More recently, 
Altshuler, et al. (Proc Natl Acad Sci USA, 107(28): 12446--12450, 2010)
claimed that AQO failed also for random instances of the
NP-complete Exact Cover problem.
In this paper, we make clear that all these negative results are only for a
{\em specific} AQO algorithm. 
We do so by demonstrating different AQO
algorithms for the same problem for which their arguments no longer
hold. Whether AQO fails or succeeds for solving the NP-complete problems (either
the worst case or the average case) requires further
investigation. Our AQO algorithms for Exact Cover and 3SAT are
based on the polynomial reductions to the NP-complete 
Maximum-weight Independent Set (MIS) problem.
\end{abstract}

\vspace*{10pt}


\section{Introduction}
A quantum computer promises extraordinary power over a classical computer,  as demonstrated by 
Shor~\cite{shor} in 1994 with the polynomial quantum algorithm for solving the factoring problem, 
for which the best known classical algorithms are exponential.
Just how much more powerful are quantum computers? 
In particular, can
a quantum computer 
 solve NP-complete problems~\cite{garey-johnson}
{\em more} efficiently than a classical computer?
NP-complete problems are the ``hardest'' problems in NP ---  in the
sense that if one
can solve one NP-complete problem efficiently (i.e., in polynomial
time) then one can solve all the problems in NP in polynomial time.
In 2000, Farhi et al.\cite{FGGS00,FGGLLP01} proposed an adiabatic quantum algorithm
as an alternative quantum paradigm to directly solve NP-hard 
optimization problems, which are polynomially equivalent to their
corresponding NP-complete decision problems.
Apparently, the same idea to the adiabatic quantum optimization,
under a different name of
 {\em quantum annealing},  was first put forward by Apolloni et al. in
 1988, 
 see \cite{ST2006,ST2008} and references therein for a history of
 the field.

\subsection{Adiabatic Quantum Algorithm}

An adiabatic quantum algorithm is described by a time-dependent system Hamiltonian
\begin{equation}
\ham(t) = (1-s(t))\ham_{\ms{init}} + s(t) \ham_{\ms{problem}}
\end{equation}
for $t \in [0,T]$, $s(0)=0$, $s(T)=1$.
There are three components of $\ham(.)$: 
(1) initial Hamiltonian: $\ham(0)=\ham_{\ms{init}}$;
(2) problem Hamiltonian:  $\ham(T)=\ham_{\ms{problem}}$;
and (3) evolution path: $s : [0,T] \longrightarrow [0,1]$, e.g., $s(t)=\frac{t}{T}$.
$\ham(t)$ is an adiabatic algorithm for a problem if we encode the problem into the problem 
Hamiltonian $\ham_{\ms{problem}}$ such that the ground state of $\ham_{\ms{problem}}$ corresponds to the answer to
the problem. The initial Hamiltonian $\ham_{\ms{init}}$ is chosen to be non-commutative with $\ham_{\ms{problem}}$
and its ground state must be known  and experimentally constructable, e.g.,  $\ham_{\ms{inital}} = -\sum_{i \in \ver(G)} \Delta_i \sigma_i^x$.
Here $T$ is the running time of the algorithm.
According to the adiabatic theorem, if $\ham(t)$ evolves ``slowly''
enough, or equivalently, if $T$ is ``large'' enough,
which scales polynomially with the inverse
of the {\em minimum spectral gap} $\gmin$ (the difference between the two lowest energy levels)
 of the system Hamiltonian, 
the system remains in the ground state of $\ham(t)$, and consequently, ground state of $\ham(T)=\ham_{\ms{problem}}$ gives the solution to the problem.

This computational model is referred as the {\em Adiabatic Quantum
Computation} (AQC). It has been shown\cite{ADKLLR04,lidar-equiv} that AQC 
is polynomially equivalent to 
 conventional quantum computation (quantum circuit model). 
For the optimization problem, the problem Hamiltonian can be
expressed as a diagonal matrix in the computational basis. 
That is, let $f_{\ms{problem}} : \{0,1\}^n \longrightarrow \reals$ be a cost function of the optimization problem such that the minimum 
of the $f_{\ms{problem}}$ corresponds to the solution of the optimization problem, 
then the corresponding problem Hamiltonian $\ham_{\ms{problem}}$ is the Hamiltonian with $f_{\ms{problem}}$ as the energy function:
namely, $\ham_{\ms{problem}} = \sum_{x \in \{0,1\}^n} f_{\ms{problem}}
(x)\ket{x}\bra{x}$ (which needs to be expressible in polynomial
resources, such as Eq. (4)
).
 Hereafter we use the cost
function and the energy function of the problem Hamiltonian
interchangeably.
 It is
worthwhile to emphasize here that given a problem, there can be many
possible cost functions and thus many possible problem Hamiltonians
for the same problem. 
This
restricted model is referred as  the {\em Adiabatic Quantum Optimization}
(AQO) (which is no longer polynomially equivalent to the quantum
circuit model). We remark that this distinction between AQC and AQO was not
made by Altshuler~et~al.~\cite{altshuler-2009} (They referred both as
AQO). 
In this paper,  the focus is on this restricted model.



\subsection{What Does ``AQO Fails'' Mean?}

The NP-complete problems that were initially proposed for AQO by Farhi et al.\cite{FGGS00,FGGLLP01} were 3SAT
and a special case of 3SAT --- Exact Cover 3 (EC3). As an example, they
proposed a clause-violation cost function as the energy function of
the problem Hamiltonian. 
Namely, $f_{\ms{problem}} (x) = \mbox{number of clauses violated by the assignment } x$. 
This cost function (with perhaps constant difference) has been adopted by almost all the other
adiabatic quantum computation works. 
In particular, van Dam and Vazirani~\cite{DV01} claimed that
AQO failed
to solve a family of 3SAT instances by showing that the AQO algorithm
with this specific clause-violation cost function
had the exponentially small minimum spectral gap. See Discussion for
more discussion on some other similar claims(\cite{DMV01,Reichardt-04}).
Recently, 
 Altshuler et al.~\cite{altshuler-2009} claimed that
AQO failed for the average case of
NP-complete problems by arguing this specific clause-violation cost
function based AQO
algorithm failed for random instances of EC3, because of the Anderson localization phenomenon.
While the cost function proposed is a ``natural'' one, nevertheless, it
is not the only possible one.
Recall that, according to the formulation of an AQO
algorithm,
the requirement of the problem Hamiltonian  
is that the ground state corresponds to the solution.
There can be other cost functions with the same minimum (solution),
e.g., see the reduction below.
That is, there are other problem Hamiltonians that have the same
ground state but a different energy spectrum. 
Given a problem, there are three components (initial Hamiltonian, problem Hamiltonian, and evolution path) that specify
an AQO algorithm for the problem.
 A change in one component (e.g. problem Hamiltonian) will result in a
 different AQO algorithm for 
the same problem. 
So how to prove that AQO fails for a problem? 
When
is it sufficient to argue for one specific cost function (and thus
specific problem Hamiltonian) and generalize to all other possible
problem Hamiltonians?


\subsection{NP-Complete Reduction Based Problem Hamiltonian}
In this paper, we make clear that the arguments in both van Dam and
Vazirani~\cite{DV01} and Altshuler~et~al.~\cite{altshuler-2009}
for their specific AQO algorithm do not generalize. We do so by
concretely describing another problem Hamiltonian for the same problem
to which their arguments 
no longer apply. 
The counter-arguments we provide are simple enough that researchers
who are not familiar with AQC and/or do not fully understand the arguments in
\cite{altshuler-2009,DV01}
can easily follow.
Our problem Hamiltonians are
based on the polynomial reductions 
-- NP-complete problems, by definition, can be polynomially reducible to each other --
to 
the Maximum-weight Independent Set problem (see e.g.~\cite{garey-johnson}), which is one of the
well studied NP-complete problems.
Perhaps, a more interesting and important question 
is: would the NP-complete reduction make a difference for the adiabatic complexity of the problem? 
Recall that
the reduction requires only the solution  to be preserved,
i.e. there is 
 a polynomial time algorithm that maps
the solution to the 
reduced problem  to  the solution to the original problem and vice versa (see e.g. \cite{DPV}).
In other words, the reduction might only preserve the solution
(i.e. the ground state) and alter the energy levels of the problem
Hamiltonian. 
In \cite{ChoiAvoid}, we demonstrate that even in small examples, 
 the minimum spectral gap can be increased drastically when the
 excited energy levels are changed,
due to the freedom in selecting parameters in the problem Hamiltonian.
Thus,
different reductions are possible, giving rise to different problem Hamiltonians,
and thus different AQO algorithms, for the same problem.
What are the time complexities of these different AQO algorithms? 
Unfortunately, at this point, the analytical analysis of
these algorithms, which requires bounding the minimum spectral gap of
the system Hamiltonian,
remains challenging. 
Whether the time complexity of each of these AQO algorithms is polynomial or
exponential remains open and requires further investigation.

\section{Methods and Results}
In the following, we first recall the NP-complete Maximum-weight Independent
Set (MIS) problem, and a problem Hamiltonian for solving MIS. Then we describe 
a simple reduction from Exact Cover to MIS that results in a different
problem Hamiltonian (and thus different
AQO algorithm) for Exact Cover. We then make clear
 that the argument in Altshuler~et~al.~\cite{altshuler-2009} does
not apply to this algorithm. Similarly, we recall the well-known
reduction from 3SAT to MIS, and point out that the argument in
\cite{DV01} does not apply to the MIS-based AQO algorithm.


\subsection{Maximum-weight Independent Set}
\label{sec:AA-mis}

Formally, the Maximum-Weight Independent Set 
problem is as follows:

\smallskip
{\bf Input:} An undirected graph $G (=(\ver(G),\edge(G)))$, where each vertex $i \in \ver(G) = \{1, \ldots, n \}$ is weighted by a
positive rational number $c_i$

{\bf Output:} A subset $S \subseteq \ver(G)$ such that
$S$ is independent (i.e., for each $i,j \in \ver(G)$, $i\neq j$, $ij
\not \in \edge(G)$) and the total
{\em weight} of $S$ ($=\sum_{i \in S}
c_i$) is maximized. 
Denote the optimal set by $\wmis(G)$.
\smallskip

This is referred as the {\em optimization} version of {\sc MIS}. The
corresponding {\em decision} problem is:

\smallskip
{\bf Question:} Is there an independent subset $S
\subseteq \ver(G)$ such that the total weight is at least $k$? (where
the positive rational number $k$ is the extra input parameter.)
\smallskip

Technically speaking, the decision version of MIS is NP-complete, and
the optimization version of MIS is NP-hard. Since the decision
problem and the optimization problem are polynomially transformable to
each other, in the following, we will simply refer to the MIS
problem, without explicitly mentioning which version (optimization or
decision). In general, which version 
should be clear from the context.

One way to solve the MIS problem is by maximizing a quadratic binary
function $\oy$
(also known as pseudo-boolean function) defined in the following
theorem (Theorem 5.1 in \cite{minor-embedding}).
\begin{thm}
If $J_{ij} \ge \min\{c_i,c_j\}$ for all $ij \in \edge(G)$, then the maximum
  value of
  \begin{equation}
\oy(x_1,\ldots, x_n) = \sum_{i \in \ver(G)}c_i x_i - \sum_{ij \in \edge(G)}
  J_{ij}x_ix_j
\label{eq:Y}
  \end{equation}
is the total weight of the MIS. 
In particular if $J_{ij} > \min\{c_i,c_j\}$ for all
      $ij \in \edge(G)$, then $\wmis(G) = \{i \in \ver(G) : x^*_i = 1\}$,
where $(x^*_1, \ldots, x^*_n) = \argmax_{(x_1, \ldots, x_n) \in \{0,1\}^n}
\oy(x_1, \ldots, x_n)$.
\label{thm:mis}
\end{thm}

Notice that in this formulation, we only require $J_{ij} > \min\{c_i,c_j\}$, and thus there is freedom in
choosing this parameter. 
It is easy to see (by changing the variables $x_i=\frac{1+s_i}{2}$) that MIS is equivalent to 
minimizing the {\em Ising energy
   function} :
\begin{eqnarray}
  \energy(s_1, \ldots, s_n) &=& \sum_{i \in \ver(G)} h_i s_i + \sum_{ij \in \edge(G)} J_{ij}s_is_j,
\end{eqnarray}
which is the 
eigenfunction of the following
 {\em Ising Hamiltonian}:
\begin{equation}
\ham_{C} = \sum_{i \in \ver(G)} h_i \sigma^z_i + \sum_{ij \in \edge(G)} J_{ij}
\sigma^z_i \sigma^z_j
\label{eq:Ising}
\end{equation}
where $h_i = \sum_{j \in \nbr(i)}
  J_{ij} - 2c_i$, $\nbr(i) =\{j: ij \in \edge(G)\}$,
for $i \in \ver(G)$.

In other words, using the Ising Hamiltonian $\ham_{C}$ as the problem
Hamiltonian, we have an AQO algorithm for solving the MIS
problem. At the end of the evolution, 
 the ground state of $\ham_{C}$, say $\ket{x_1^*x_2^*\ldots x_n^*}$,
 corresponds to the maximum-weight independent set, namely $\wmis(G) =
  \{i: x_i^* = 0\}$.


\subsection{Exact Cover}
Formally, the \EC is as follows:

{\bf Input:} A set of $m$ elements, $X = \{c_1, c_2, \ldots, c_m\}$;  
a family of $n$ subsets of $X$, $\mathcal{S} = \{S_1, S_2, \ldots, S_n\}$, where $S_i \subset X$

{\bf Question:} Is there a subset $I \subseteq \{1, \ldots, n\}$ 
such that $\cup_{i \in I} S_i = X$, where $S_i \cap S_j = \emptyset$ for $i\neq j \in I$? 
Here $\{S_i: i \in I\}$ is called an {\em exact cover} of $X$.

\begin{figure}[ht]
\centerline{\includegraphics[width=0.6\textwidth]{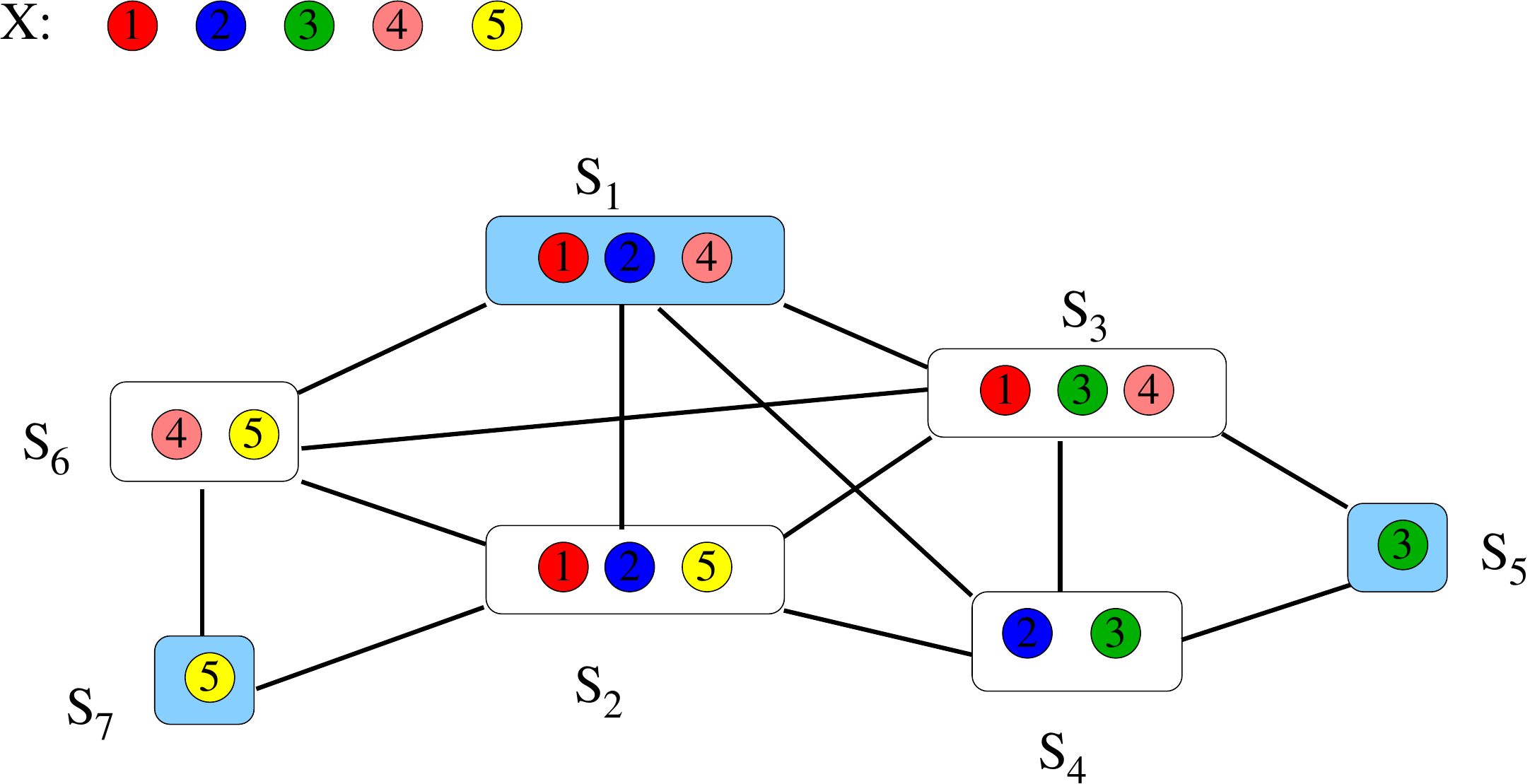}}
\caption{$X=\{c_1,c_2,c_3,c_4,c_5\}$, and $\mathcal{S} = \{S_1, S_2, \ldots, S_7\}$, with 
$S_1=\{c_1,c_2,c_4\}$, $S_2=\{c_1,c_2,c_5\}$, $S_3=\{c_1,c_3, c_4\}$,$S_4 = \{c_2, c_3\}$, $S_5=\{c_3\}$, $S_6=\{c_4,c_5\}$, $S_7=\{c_5\}$.
Here $\{S_1, S_5, S_7\}$ is the exact cover of $X$.
In this example, each element $c_i$ appears in exactly three subsets. For instance, element $c_1$ (the red ball) appears in 
$S_1, S_2$ and $S_3$.
(a) The problem instance can be viewed as a graph 
 $G_M$ with $n$ vertices, where vertex $i$ corresponds to the set
 $S_i$, and there is an edge between two vertices if and only if $S_i$
 and $S_j$ overlap (share a common element). 
There is an exact cover of $X$ if and only if  the weight of $\wmis(G_M)$  is $m$, where the weight of vertex $i$ is the number of elements in $S_i$.
(b) The problem can also be reduced to the following instance of positive 1-in-3SAT:
$\Psi(x_1, \ldots, x_7) = C_1 \wedge C_2 \wedge C_3 \wedge C_4 \wedge C_5$ where
 $C_1 = x_1 \vee x_2 \vee x_3$, $C_2 = x_1 \vee x_2 \vee x_4$, $C_3 = x_3 \vee x_4 \vee x_5$,
    $C_4 = x_1 \vee x_3 \vee x_6$, $C_5 = x_2 \vee x_6 \vee x_7$.
There is an exact cover of $X$ if only if $\Psi$ is satisfiable in that there is exactly one variable in each clause is true.
}  
\label{fig:EC3-example}
\end{figure}

See Figure~\ref{fig:EC3-example} for an example. In particular, in this example, each element $c_i \in X$ appears exactly in three subsets. This kind of special instance is referred as
EC3, which can then be polynomially reducible ($\le_{P}$) to the positive 1-in-3SAT problem. 

\paragraph{\SEC $\le_{P}$ positive 1-in-3SAT.}
Given an instance of \SEC with an $m$-element set $X$ and $n$ subsets $S_1, \ldots, S_n$,
we construct a 3CNF boolean formula $\Psi(x_1, \ldots, x_n) = C_1 \wedge \ldots \wedge C_m$ with $n$ variables and $m$ clauses.
For each element $c_i \in X$, let $S_{i_1}$, $S_{i_2}$, $S_{i_3}$ be the three sets that consist of $c_i$. By definition, exactly one of these three sets needs to be in the exact cover of $X$. Therefore, if we define a binary variable $x_k$ for each set $S_k$ such that $x_k=1$ if $S_k$ is in the exact cover, and $x_k=0$ otherwise, we require exactly one variable in $C_i = x_{i_1} \vee x_{i_2} \vee x_{i_3}$ to be satisfied. That is, there is an exact cover to the original problem if and only if the formula 
$\Psi(x_1, \ldots, x_n) = C_1 \wedge \ldots \wedge C_m$ is satisfiable in that there is exactly one variable in each clause is satisfied, which is the positive 1-in-3SAT problem.
See Figure~\ref{fig:EC3-example}(b) for an example of the reduction.

One way of solving the positive 1-in-3SAT problem is to minimize the
following clause-violation
 cost function 
$$
\eng_\Psi(x_1, \ldots, x_n) = \sum_{i=1}^{m} (x_{i_1} + x_{i_2} + x_{i_3} -1 )^2.
$$
$\Psi$ is satisfiable if and only if the minimum of $\eng_\Psi$ is $zero$ (i.e. no violation).
The corresponding problem Hamiltonian based on this cost function as
used by  
Altshuler~et~al.\cite{altshuler-2009} (and Young~et~al.~\cite{young-2009})\footnote{The sign of $\sigma_i^z$ term is in opposite because they use $x_i=\frac{1-s_i}{2}$ instead. $I_{ij}$ was called $J_{ij} (= \frac{1}{2}(J_{ij} + J_{ji}))$ in \cite{altshuler-2009}.} is
\begin{eqnarray}
  \ham_{A} = \sum_{i \in \ver(\GEC)}B_i \sigma_i^z +  \sum_{ij \in \edge(\GEC)} I_{ij} \sigma_i^z \sigma_j^z
\end{eqnarray}
where $B_i$ is the number of clauses that contains variable $x_i$,  and $I_{ij}$ 
 is the number of clauses that contains both $x_i$ and $x_j$, and  
$\ver(\GEC) = \{1,  \ldots, n\}$,
 and $\edge(\GEC) = \{ij: x_i \mbox{ and } x_j \mbox{ appear in a clause.}\}$.

Next, we show that there is another way of looking at the Exact Cover
problem, which leads to a simple polynomial reduction from Exact Cover to MIS.
\paragraph{\EC $\le_{P}$ MIS.} Given an instance of \EC with an
$m$-element set $X$ and $n$ subsets $S_1, \ldots, S_n$.
Recall that each element in $X$ can appear in exactly in one set. If
two sets overlap, e.g. $S_1$ and $S_2$ in
Figure~\ref{fig:EC3-example}, then they can not both appear in the
exact cover. Therefore, 
 if we construct a graph $G_M$ with $n$ vertices, where vertex $i$
 corresponds to the set $S_i$, and 
there is an edge between two vertices if  $S_i$ and $S_j$ overlap.
Then an exact cover of $X$ will correspond to an independent set of  $G_M$. 
Moreover, if we let  the weight of vertex $i$ be the number
 of elements in $S_i$,
then
there is an exact cover to the original problem if and only if  the weight
of $\wmis(G_M)$  is $m$.

For EC3, it is easy to see that $\GEC$ and $G_M$ are exactly the same because  there is one-one corresponding between the variable $x_i$ and the set $S_i$ ($\ver(\GEC)=\ver(G_M)$), and  ``$x_i \mbox{ and } x_j$ appear in a  clause'' is equivalent to 
 ``$S_i$ and $S_j$ share a common element'' ($\edge(\GEC)=\edge(G_M)$).
Based on this reduction, we therefore have the following problem
Hamiltonian for the same problem:
{\small
\begin{eqnarray}
\ham_C= \sum_{i \in \ver(\GEC)} \left(\sum_{j \in \nbr(i)}J_{ij}-2B_i\right)\sigma^z_i + \sum_{ij \in \edge(\GEC)} J_{ij} 
\sigma^z_i \sigma^z_j
\end{eqnarray}
}
where $J_{ij} > \min\{B_i,B_j\}$.

\paragraph{Comparison of $\ham_{A}$ and $\ham_C$.}
For comparison, recall that 
$2B_i = \sum_{j \in \nbr(i)} I_{ij}$, let us write
$J_{ij} = 2 I_{ij} + D_{ij}$, where $D_{ij} >\min\{B_i,B_j\} - 2I_{ij}$.
Then we have 
$$\ham_C= 2\ham_{A} + \sum_{i \in \ver(\GEC)}  \sum_{j \in \nbr(i)}D_{ij} \sigma^z_i + \sum_{ij \in \edge(\GEC)} D_{ij} 
\sigma^z_i \sigma^z_j.$$ 
Alternatively, as we show in \cite{Choi-PNAS}, for $D_{ij}>0$, we can view $\ham_C$ as obtained from
the following clause-violation cost function:

\begin{eqnarray}
  \label{eq:1}
\eng'_\Psi(x_1, \ldots, x_n) 
&=& \sum_{k=1}^{m} (x_{k_1} + x_{k_2} +
x_{k_3} -1 )^2 + \sum_{ij \in \edge(\GEC)} D_{ij} x_i x_j\\
&=&\eng_\Psi(x_1, \ldots, x_n) + \sum_{ij \in \edge(\GEC)} D_{ij} x_i x_j
\end{eqnarray}

Notice that the extra term ($\sum_{ij \in \edge(\GEC)} D_{ij} x_i x_j$)  will not contribute to the energy function for the truth assignment because at least one of 
$\{x_i, x_j\}$ (for $ij \in \edge(\GEC)$)   will be zero. 
In other words, both $\eng_\Psi$ and $\eng'_\Psi$ are the cost
functions for the same problem instance $\Psi$, 
where $D_{ij}>0$ can be arbitrary. 
The question is: does this extra term (for arbitrarily chosen
$D_{ij}>0$)  matter to the performance of the adiabatic algorithm?

In \cite{altshuler-2009}, Altshuler et al. claimed that the AQO algorithm  with problem Hamiltonian $\ham_{A}$ failed
with high probability for randomly generated instances of EC3 due to the
Anderson localization(AL). 
The authors claimed that 
the correctness of their argument did not rely on the specific form of 
the problem Hamiltonian for EC3, but only depended on the properties of the problem instance $B_i$ and $I_{ij}$.
However, our problem Hamiltonian  $\ham_{C}$ belies their claim.
Their argument necessarily depends on the energy function of the
problem Hamiltonian.
While the energy function for $\ham_{A}$ only depends on $B_i$ and $I_{ij}$, the energy function for $\ham_C$
also depends on the extra $D_{ij}$ whose values have a range to
choose. Therefore, their argument which depends on the random property
of parameters does not apply to  the algorithm based on $\ham_{C}$ because $D_{ij}$ are not random.
Perhaps, their main message is that from the
physics point of view, if AL occurs, then the AQO algorithm will
require exponential time, as shown by Amin and Choi in~\cite{AC09}.
So what are the necessary conditions for AL to  occur?
Can the arguments in  \cite{altshuler-2009} {\em be modified} such that AL
still occurs, and the AQO algorithm with $\ham_{C}$ for arbitrarily
chosen $D_{ij}$ will also fail?
Here we emphasize that while the connectivity of the graph is random,
the values of $D_{ij}$ (at least $\min\{B_i,B_j\} -
2I_{ij}$) are free to choose. To show that the AQO algorithm with
$\ham_{C}$ fails, one would need to show that the algorithms
correspond to
{\em all} possible values of $D_{ij}$ {\em all} fail.
Or alternatively, can one show that there does not exist $D_{ij}$ (not
necessarily efficiently computable) such that the corresponding
adiabatic algorithm is not at least exponential?
In \cite{young-2009}, Young~et~al. used QMC to show that the minimum
spectral gap $\gmin$ of the AQO algorithm (with the same problem Hamiltonian $\ham_{A}$) is exponentially small. 
It will be interesting to see  the $\gmin$ result (for the same set of
instances) using this new problem Hamiltonian $\ham_{C}$ for some
$D_{ij}$.

\subsection{3SAT}
\begin{figure}[h]
  \centering
  \includegraphics[width=0.6\textwidth]{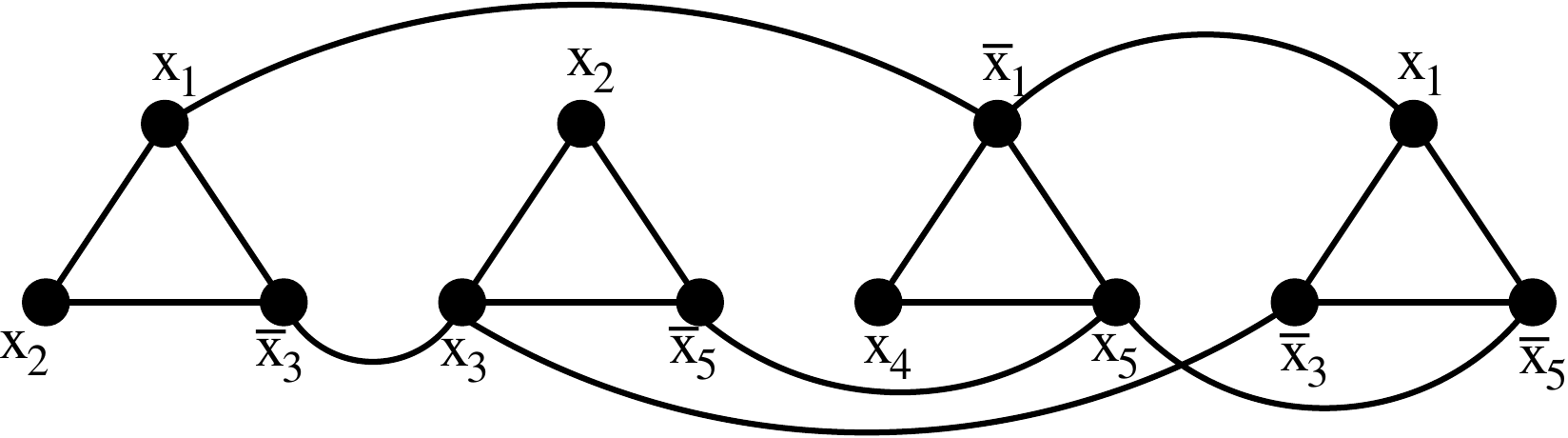}
  \caption{Graph $\GSAT$ corresponds to $\Psi(x_1, \ldots, x_5) =
    (x_1 \vee x_2 \vee \bar{x}_3) \wedge (x_2 \vee x_3 \vee \bar{x}_5)
    \wedge (\bar{x}_1 \vee x_4 \vee x_5) \wedge (x_1 \vee
    \bar{x}_3 \vee \bar{x}_5)$. $\Psi$ is satisfiable if and only if $\GSAT$ has a MIS of size $4$.
}
  \label{fig:SAT-MIS}
\end{figure}

Similarly, for 3SAT, there is a well-known reduction to MIS (which is one of the first NP-complete reductions, to show the NP-hardness of MIS)~\cite{garey-johnson}. For completeness, here we recall the reduction:

\paragraph{\SAT $\le_{P}$ MIS.} Given a \SAT instance $\Psi(x_1, \ldots, x_n) = C_1 \wedge \ldots \wedge C_m$ with $n$ variables and $m$ clauses, we construct a (unweighted) graph $\GSAT$ as follows:
\begin{itemize}
\item For each clause $C_i = y_{i_1} \vee y_{i_2} \vee y_{i_3}$, we construct a triangle with three vertices labeled  
accordingly, i.e., with $y_{i_1}, y_{i_2}, y_{i_3}$, where $y_j \in \{x_j, \bar{x}_j\}$. Therefore,  $\GSAT$ consists of $3m$ vertices.\item There is an edge between two vertices in different triangles if there labels are in conflict. That is, for $i \neq j$,   $i_sj_t \in \edge(\GSAT)$ if and only if $y_{i_s} = \bar{y}_{j_t}$. 
\end{itemize}
One can then show that $\Psi$ is satisfiable if and only if $\GSAT$ has a MIS of size $m$.
See Figure~\ref{fig:SAT-MIS} for an example.

In \cite{DV01}, van Dam and Vazirani constructed a special set of \SAT instances.
They showed that  the clause-violation cost function based
AQO algorithm 
would take exponential time for these instances  because of the exponential small $\gmin$.
Farhi~et~al.\cite{diff-path1} showed that the exponential small gap could be overcome by different initial Hamiltonians. 
Here we point out also that the argument in \cite{DV01} does not apply to the MIS based problem Hamiltonian. 
In order to show that the exponential small gap, their argument requires the cost function to be ``$\epsilon$-deceptive monotone''. 
For completeness, here we recall the definitions:

\begin{mdef} 
(Definition 1 in \cite{DV01})
Consider the hypercube $\{0,1\}^n$ with the partial ordering on its
strings $x \preceq y$ if and only if $x_i \le y_i$ for all $i=1,
\ldots, n$. A function $f: \{0,1\}^n \longrightarrow \reals$ is
monotonically decreasing if $x \preceq y$ implies $f(x) \ge
f(y)$. Similarly for monotonically increasing functions.
\end{mdef}

\begin{mdef}
(Definition 2 in \cite{DV01})
  A cost function $f: \{0,1\}^n \longrightarrow \reals$ is
  ``$\epsilon$-deceptive monotone'' if it is monotone for all but a
  fraction $\epsilon$ vertices comprising the top layers of the hypercube.
\end{mdef}

However, it is easy to see that the cost function (i.e. the pseudo-boolean function $\oy$ in Theorem~\ref{thm:mis}) for MIS
is not monotone. For example, $\oy(1,0,0, \ldots, 0) > \oy(1,1,0,
\ldots, 0)$ where  vertex $1$ and vertex $2$ are adjacent, 
while $\oy(1,0,0, \ldots, 0) < \oy(1,0,0,
\ldots, 1)$ where vertex $1$ and vertex $n$ are independent.
Therefore their argument no longer applies to this cost function, and
one can no longer conclude that this AQO algorithm also fails for the instances they
constructed. 


\section{Discussion}
\label{sec:discussion}
AQO was originally proposed in \cite{FGGS00,FGGLLP01}  as an energy minimization algorithm
 that aims to use coherent quantum evolution to avoid trapping 
in the local minima 
that trip classical algorithms of NP-hard optimization problems.
Early on, van Dam et al.~\cite{DV01,DMV01} and
Reichardt~\cite{Reichardt-04} showed that the AQO
 algorithm failed to avoid local minima and would take exponential time for
computing the minimum of some cost function $f$, where their AQO
algorithm is defined such that $f$ is the energy function of the problem
Hamiltonian. In other words, in their formulation, the energy function
of the problem
Hamiltonian of their AQO algorithm is same as the cost function of the
problem. However,
given a problem, there are in general many possible cost functions
with the same minimum. 
Thus, their
AQO algorithm is just a {\em specific} algorithm of our more general AQO
definition. 
This perhaps rather obvious
 difference
unfortunately is not clear to even quantum computation experts who
are not working on the adiabatic quantum computation, and the results
of van Dam et al.\cite{DV01,DMV01} and Reichardt\cite{Reichardt-04} were
widely believed that they held for all AQO algorithms.
Technically speaking, the lower bound of this specific AQO
algorithm for computing the minimum of $f$ does not constitute a lower
bound for all AQO
 algorithms of the same problem. 
Nevertheless, do these results provide
``convincing evidence'' that AQO would fail to solve problems with
many local minima?
For this purpose, we constructed a family of MIS graphs in which there
are exponentially many local minima, our initial results, which were 
explained by the first order quantum phase transition~\cite{AC09},
agreed with the speculation -- that is, the system got trapped in the
local minima and the particular AQO algorithm failed.
However, in \cite{ChoiAvoid}, we showed
that the exponential small gap (caused by the system getting trapped
in the local minima) could be overcome by changing the
parameters of the problem Hamiltonian without changing the problem
(that is, the global minimum and the exponential local minima remain
the same). Although the result is only numerical and supported by visualization,
this small example serves to  clarify that
it is not sufficient to consider one specific problem Hamiltonian
(and thus specific AQO algorithm) for
proving the adiabatic lower bound of a problem.

After the appearance of the arXiv version of this paper, F.  Krzakala pointed
out to the author that the failure of ``specific'' AQO algorithms for
the random instances of some NP-complete problems was discussed in
their work \cite{Jorg1,Jorg2}.
Note that also, some initial Hamiltonian, such as the
projection Hamiltonian used in \cite{znidaric-2006}, can make the
corresponding AQO algorithm fail as shown
in~\cite{farhi-2005}. Recently, it was also shown in
\cite{farhi-2010} that the
lack of structure in the cost function of the problem Hamiltonian can
 make the corresponding AQO fail. 
Farhi et al.~\cite{farhi-2009} suggested that
an adiabatic quantum algorithm should be run on each instance with
many different random paths (as an integral part of AQO). 

In this paper, we describe different AQO algorithms for the
NP-complete Exact Cover and 3SAT problems. These algorithms are rather
straightforward, from the computer science point of view, as they are
based on the simple NP-complete reductions. However, they serve to
further clarify the distinction between one specific problem Hamiltonian AQO
algorithm and general AQO algorithms for a problem.
In particular, we make clear that the arguments in both van Dam and
Vazirani~\cite{DV01} and Altshuler~et~al.~\cite{altshuler-2009}
for their specific AQO algorithm
do not apply to the AQO algorithms we
describe here. 
While from the physics point of view, ``Anderson Localization'' makes an 
AQO algorithm fail, does AL necessary occur for all
AQO algorithms of the same problem? What are the assumptions or
conditions for AL to occur? 
In particular, can one modify the arguments in
Altshuler~et~al.~\cite{altshuler-2009} such that the AQO algorithm
described here also fails?
Whether the time complexity of these different
AQO algorithms is  exponential or not requires more rigorous analysis.
Our counter-arguments are so simple that one does not need to 
fully understand the arguments in \cite{altshuler-2009,DV01} nor to  have
advanced algorithmic knowledge in order to understand.  This perhaps highlights the
challenge of this interdisciplinary research, and calls for more
serious investigation.

\section*{Acknowledgements}
I would like to thank Peter Young and Boris Altshuler for the
 discussion. 
I would like to thank Siyuan Han, 
Mohammad Amin, Neil Dickson, Robert Raussendorf, Tzu-Chieh Wei and
Pradeep Kiran for their comments. 
I would like to thank my very enthusiastic students in my adiabatic quantum computing class: 
Ryan Blace,
Russell Brasser, 
Mark Everline,
    Eric Franklin,
     Nabil Al Ramli,
  and   Aiman Shabsigh.
Thanks also go to David Sankoff and David Kirkpatrick for the encouragement.

\bibliographystyle{abbrv}


\end{document}